\documentclass[a4paper]{jpconf}
\usepackage{graphicx}
\begin{document}

\title{IceCube Search for Galactic Neutrino Sources based on Very High Energy $\gamma$-ray Observations}

\author{Ali Kheirandish for the IceCube \& HAWC Collaborations}

\address{Department of Physics \& Wisconsin IceCube Particle Astrophysics Center, University of Wisconsin, Madison, WI 53706, USA}

\ead{akheirandish@icecube.wisc.edu}

\begin{abstract}
Galactic cosmic rays reach energies of at least several PeV, and their interactions should generate $\gamma$-rays and neutrinos from decay of secondary pions. Therefore, Galactic sources have a guaranteed contribution to the total high-energy cosmic neutrino flux observed by IceCube.  Assuming that the highest energy $\gamma$-rays are pionic, promising neutrino source candidates have been identified based on their spectra, and observing them is likely over the lifetime of the IceCube experiment. Here, we present the search for Galactic sources of high-energy cosmic neutrinos by focusing on 
sources identified by HAWC's very high energy $\gamma$-ray survey.
\end{abstract}

\section{Introduction}\label{sec:intro}
In the search for the origin of very high energy particles, the Galactic plane provides a rich environment to study the nature and mechanism of particle acceleration. Galactic cosmic ray accelerators are believed to produce cosmic rays (CRs) up to several PeV, the "knee" in the CR spectrum. The CRs interact with the dense environment in the Milky Way, and produce charged and neutral pions, which subsequently decay to high-energy neutrinos and $\gamma$-rays.

Simultaneous production of neutrinos and $\gamma$-rays opens up the opportunity for multimessenger searches. 
Multimessenger searches are essential for identifying these sources since the detection of a source in $\gamma$-rays alone cannot reveal whether hadrons are being accelerated to very high energy. This is because very high energy $\gamma$-rays can also originate from the acceleration of electrons. In contrast, high-energy neutrinos can only be generated where hadrons are present. Therefore, a high-energy neutrino observation of a source will provide a smoking gun for the identification of a Galactic CR accelerator.

The very high energy $\gamma$-ray survey of the Galaxy by the Milagro Collaboration \cite{Abdo:2006fq} revealed the brightest $\gamma$-ray sources in the Northern Sky. Promising sources were identified in the Milagro sky map based on their spectra and early estimations implied their likely observation within the initial operation years of IceCube \cite{Halzen:2008zj,GonzalezGarcia:2009jc}. Additional observations by Milagro and Imaging Air Cherenkov Telescopes provided more insight on the high-energy $\gamma$-ray emission from the Milky Way. However, the prospect for the observation of these sources in IceCube became less clear with the discrepancy in the $\gamma$-ray flux and the extension of these sources reported by different observatories \cite{Gonzalez-Garcia:2013iha, Halzen:2016seh}. This necessitated a deeper survey of the $\gamma$-ray sky above 10 TeV which is now being provided by the High Altitude Water Cherenkov (HAWC) Observatory \cite{Abeysekara:2017hyn}.
\section{IceCube \& HAWC Synergy}\label{sec:ic-hawc}

The IceCube Neutrino Observatory \cite{Aartsen:2016nxy} is a high-energy neutrino detector located at the geographic South Pole. The detector consists of an array of 86 vertical strings instrumenting 1 km$^3$ of ice with optical sensors. % 
High-energy muon neutrinos interact in the ice produce relativistic muons that may travel many kilometers, creating a track-like distribution of Cherenkov photons recorded when they pass through the array. These events range in energy from 0.2 TeV to 1 EeV and allow for the reconstruction of the original neutrino direction with a median angular uncertainty of 0.5$^\circ$ for a neutrino energy of around 30 TeV. The field-of-view of IceCube spans the entire sky. % 
However, its best sensitivity for resolving sources lies in the Northern hemisphere where charged particle backgrounds, which are created locally by CRs interacting in the atmosphere, are completely removed by absorption in the Earth.

The HAWC Observatory is a TeV $\gamma$-ray observatory that is ideally suited to measure sources that may also be identifiable by IceCube. First, its large %
 field-of-view provides a survey of the Northern Sky where backgrounds in IceCube are lowest. Second, the broad energy coverage of HAWC  from 0.3 TeV to 100 TeV closely overlaps with the energy range of IceCube. %
\section{Searching for Neutrino Emission}\label{sec:ic} 
We search for neutrino emission associated with the very high energy $\gamma$-rays observed by HAWC. Assuming $\gamma$-rays detected by HAWC are pionic, high-energy neutrinos should inevitably accompany them. %
In this study, we use eight years of muon neutrino track events from the Northern Sky  \cite{Aartsen:2018ywr}.
%,
We incorporate a stacking likelihood method (see \cite{Achterberg:2006ik} for details) and test five different hypotheses for the correlation of $\gamma$-rays and neutrinos. 

The first observations of the Galactic plane using the fully realized HAWC detector were published in the 2HWC catalog \cite{Abeysekara:2017hyn}. %
While a subset of these sources are firmly identified as pulsar wind nebulae (PWN) %
, most have no firm identification. %
{ 
This joint search focuses on the set of HAWC sources from the 2HWC catalog with emission that cannot be explained entirely by a PWN counterpart.} 
Here we use an updated sky map from the catalog \cite{Abeysekara:2017hyn} with 1128 days of data to search for the Galactic sources of cosmic neutrinos.

\begin{figure}[ht!]
\centering
\includegraphics[width=5.8in]{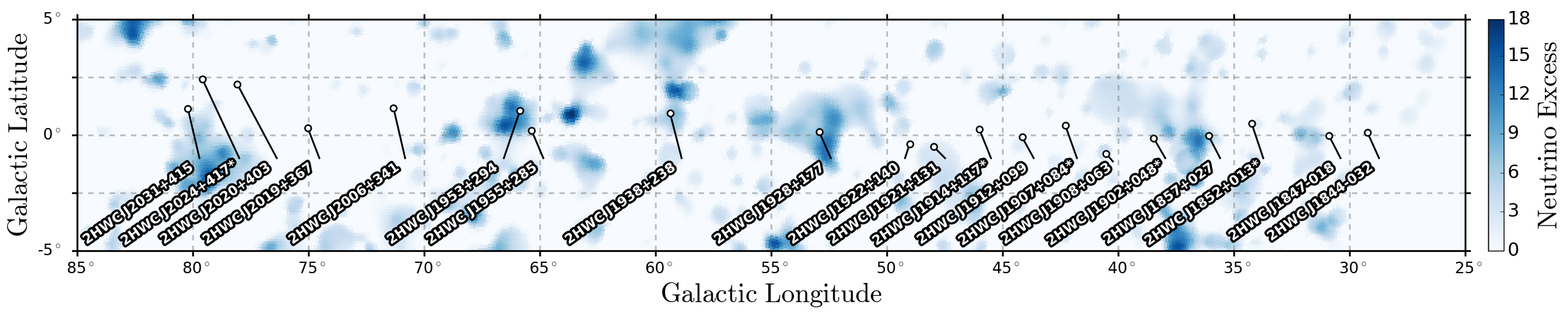}
\caption{Map of the neutrino excess in IceCube with HAWC sources overlaid.}
\label{fig:stacked_overlay}
\end{figure}

The first search is a stacking search for neutrino emission from HAWC sources that are not identified as PWN. We have excluded PWN in this search for two reasons: First, high-energy emission from PWN is generally understood to be leptonic. Second, a separate dedicated search by the IceCube collaboration has examined neutrino emission from TeV PWN \cite{qliu2019}. This work focuses on 20 sources shown in Fig. \ref{fig:stacked_overlay}. 

\begin{figure}[ht!]
\centering
\includegraphics[width=5.8in]{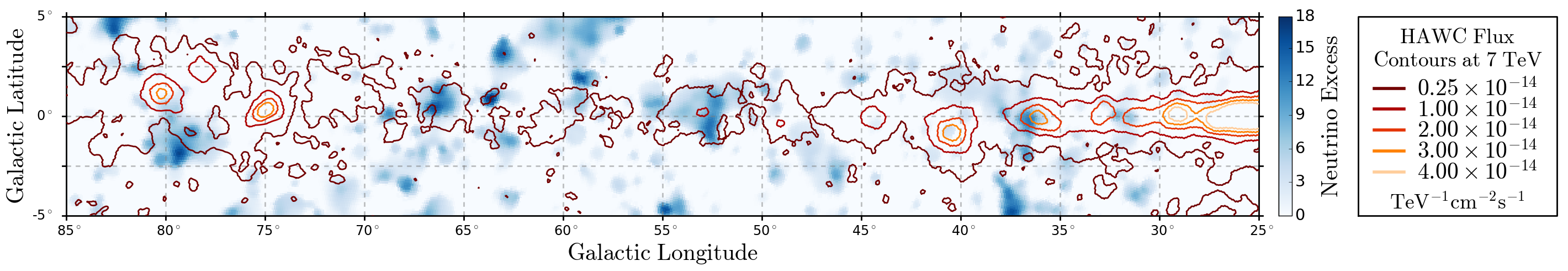}
\caption{Map of the neutrino excess in IceCube with HAWC flux contours overlaid.}
\label{fig:flux_contour_overlay}
\end{figure}

For the four other searches, we use the morphology of the $\gamma$-ray emission as reported by the HAWC Collaboration. For this purpose, we incorporate the high-energy $\gamma$-ray flux morphology as shown in Fig. \ref{fig:flux_contour_overlay} and use the $\gamma$-ray emission at each point to weight the stacking likelihood. We do this search by considering the whole plane seen by the Northern Sky muon neutrino sample (Dec. $> -5^\circ$) as well as for three regions defined {\em a priori}{: the Cygnus region, and the areas surrounding 2HWC J1908+06, and 2HWC J1857+027. These are star forming regions with high levels of $\gamma$-ray activity and young stars, and have been identified as potential neutrino emitters \cite{Halzen:2016seh,Beacom:2007yu}}.

The results of these likelihood tests are summarized in Table \ref{tab:results}. The observed excess of signal neutrinos found in these tests is not statistically significant for claiming evidence of neutrino emission associated with the $\gamma$-rays emission measured by HAWC. The most significant test corresponds to the region surrounding 2HWC J1857+027 and finds an excess of 36.7 neutrinos with a pre-trial p-value of 2\%. 

\begin{table}[h]
\caption{Results of the five analyses testing the correlation of high-energy neutrinos with $\gamma$-rays observed by HAWC. The source hypothesis, the best fit number of signal neutrinos, sensitivity and 90\% C.L. upper limit flux at 7 TeV are listed. The final column displays the pre-trial p-value found in each test.}
\label{tab:results}
 
\begin{center}
\lineup
\begin{tabular}{*{5}{l}}
\br                              
     & Best Fit  & Sensitivity & Upper Limit (90\% C.L.) & \cr
     Search & $n_s$ & $10^{- 13}$ [TeV$^{- 1}$cm$^{- 2}$s$^{-1}$] & $10^{- 13}$ [TeV$^{- 1}$cm$^{- 2}$s$^{-1}$] & p-value \cr 
\mr
Stacking & 15.4 & 0.7 & 1.5 & 0.09\cr
Northern Plane & 77.8 & 2.5 & 5.7  & 0.06\cr 
Cygnus Region & 0.0 & 1.0 & 0.4  & 0.80\cr 
J1908+063 Region & 12.0 & 0.7 & 1.3 & 0.14\cr 
J1857+027 Region & 36.7 & 0.8  & 2.1 & 0.02\cr
\br
\end{tabular}
\end{center}
\end{table}

\begin{figure}[ht!]
\centering
\includegraphics[width=0.494\linewidth]{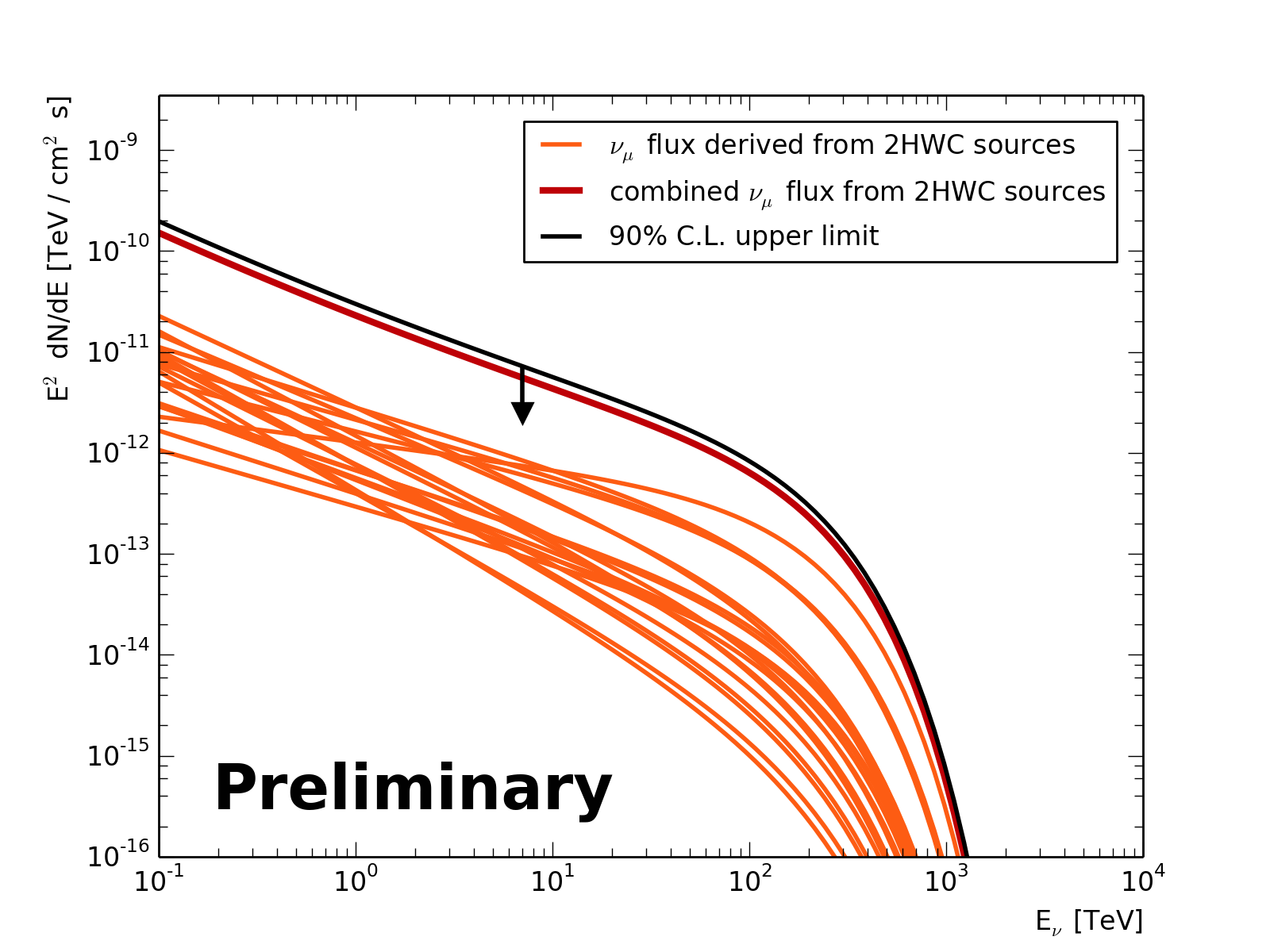}
\includegraphics[width=0.494\linewidth]{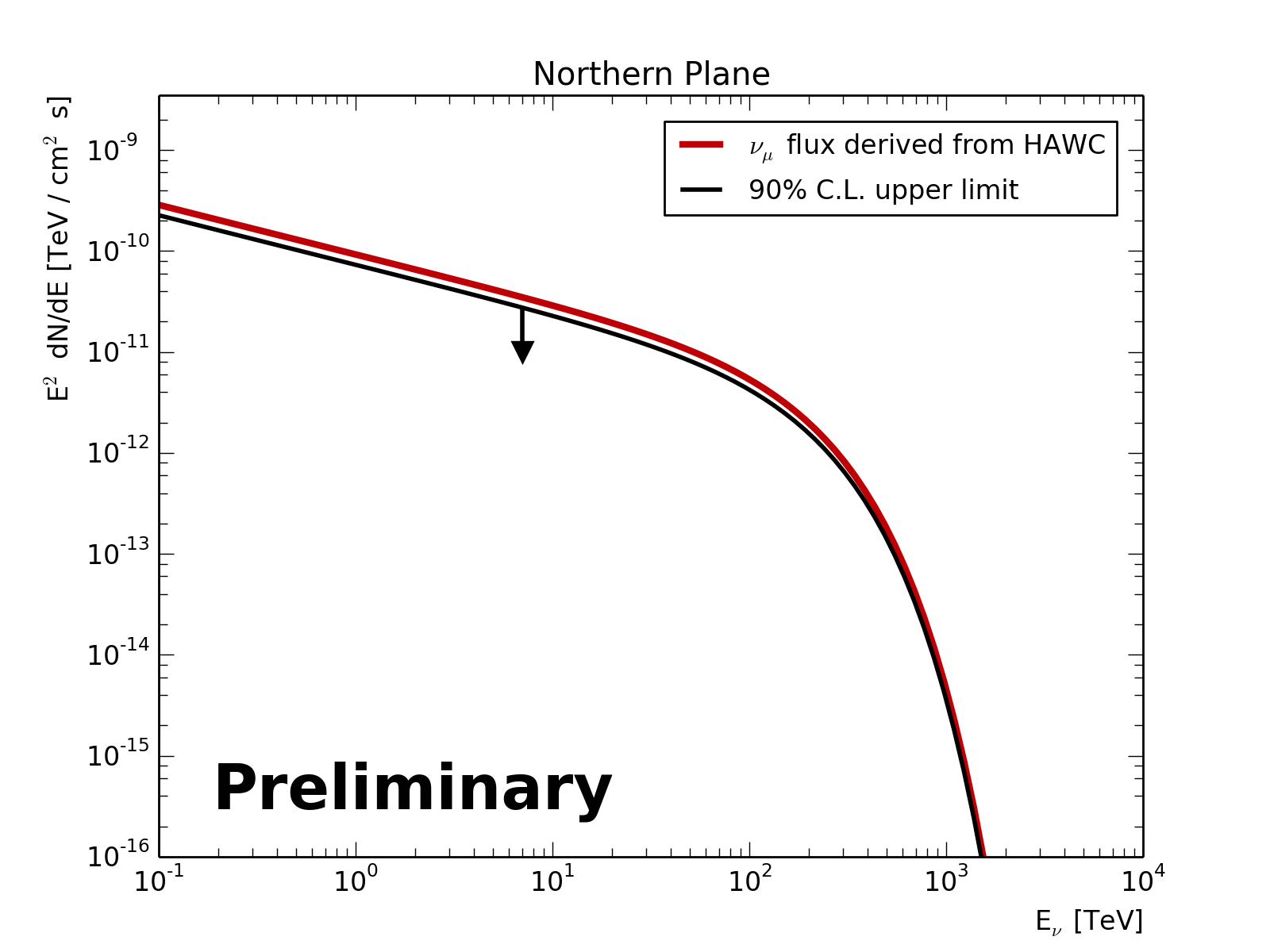}
\caption{ Left: Upper limit (90\% C.L.) on the flux of muon neutrinos (black) for the stacking search of non-PWN sources in the 2HWC catalog.The projected muon neutrino fluxes (thin orange) represent the expected flux from each source assuming that the high-energy gamma ray flux measured by HAWC is from hadronuclear interactions. The combined flux (red) shows sum of the individual fluxes. Right: Upper limit (90\% C.L.) on the flux of muon neutrinos (black) for the Northern plane and the expected neutrino flux assuming hadronuclear interactions (orange).
}
\label{fig:ul_stacked_plane}
\end{figure}

We obtain upper limits on the muon neutrino flux for each hypothesis. 
Fig. \ref{fig:ul_stacked_plane} summarizes the flux upper limits for the stacked analysis and the template search for the outer Galaxy visible in the Northern Sky. 
We project the equivalent neutrino flux of HAWC's measured $\gamma$-ray flux, assuming that all $\gamma$-rays are hadronic. The neutrino flux is derived from the $\gamma$-ray spectra measured by HAWC following \cite{Kappes:2006fg}, assuming hadronuclear interaction at the sources.

The most stringent limit is set for neutrino emission from the Cygnus region. Our search found an under fluctuation of background from this region. The upper limits on the two other regions, however, do not constrain a possible hadronic component. While the upper limit for the region surrounding 2HWC J1908+06 is marginal to the maximal hadronic $\gamma$-ray emission from the source assuming hadronic origin of the entire $\gamma$-ray flux. The excess found for the 2HWC J1857+027 region yields an upper limit on the neutrino emission much larger than the corresponding $\gamma$-ray emission. 

\section{Summary}\label{sec:sum}

Galactic accelerators have long been though to contribute to the flux of CRs arriving at Earth. The high-density regions in the Milky Way provide targets for CRs from these sources to interact and produce high-energy neutrinos and $\gamma$-rays.  
The rationale for identifying Galactic sources of cosmic neutrinos in this work assumes a close correspondence between %
  neutrinos and $\gamma$-rays.

The HAWC $\gamma$-ray observatory has surveyed the Galaxy at the highest photon energies to date. Here, we used this survey to search for the Galactic origins of the high-energy neutrinos. %

 The results reported here are marginal to the expectation, and in some cases constrain the hadronic flux component. In this study, we examined the possible neutrino emission from the sources of very high energy $\gamma$-rays observed by HAWC. This search, together with the search for neutrino emission from TeV PWN \cite{qliu2019}, provides a comprehensive study for neutrino emission from major high-energy $\gamma$-ray emitters in the Milky Way. An upcoming IceCube search for neutrino emission from binary sources will address the contribution of another potential site of CR acceleration in the Galaxy.

\section*{References}
\bibliographystyle{iopart-num}
\bibliography{references}

\end{document}